\documentclass[aps,prb,twocolumn, reprint,groupedaddress,amsmath,amssymb]{revtex4-1} 

\usepackage{graphicx}
\usepackage{bm}
\usepackage{natbib} 
\usepackage[left=.6in, right=.4in, top = .65in, bottom = 1.1in]{geometry} 

\usepackage{float}
\begin{document}


\title{Thermal transport size effects in silicon membranes featuring nanopillars as local resonators}


\author{Hossein Honarvar}
\author{Lina Yang} 
\altaffiliation[Current affiliation: ]{Department of Mechanical and Civil Engineering, California Institute of Technology, Pasadena, California 91125, USA}
\author{Mahmoud I. Hussein}%
\email{mih@colorado.edu}

\affiliation{Department of Aerospace Engineering Sciences, University of Colorado Boulder, Boulder, Colorado 80309, USA}


\date{\today}

\begin{abstract}
Silicon membranes patterned by nanometer-scale pillars standing on the surface provide a practical platform for thermal conductivity reduction by resonance hybridization. Using molecular simulations, we investigate the effect of nanopillar size, unit-cell size, and finite-structure size on the net capacity of the local resonators in reducing the thermal conductivity of the base membrane. The results indicate that the thermal conductivity reduction increases as the ratio of the volumetric size of a unit nanopillar to that of the base membrane is increased, and the intensity of this reduction varies with unit-cell size at a rate dependent on the volumetric ratio. Considering sample size, the resonance-induced thermal conductivity drop is shown to increase slightly with the number of unit cells until it would eventually level off.      
\end{abstract}


\maketitle

In semiconducting materials, heat is carried mostly by phonons which are quanta of lattice vibrations~[\onlinecite{Kittel_1976}].~This provides an opportunity to introduce significant changes to the thermal transport properties by direct engineering of the phonon characteristics$-$which are shaped primarily by the phonon band structure and the nature of the underlying scattering mechanisms~[\onlinecite{Chen_2000,*Balandin_2005}].~Recent reviews survey developments in theory, computation, and experiment pertaining to nanoscale thermal transport in a variety of materials and point to the remarkable possibilities for using nanostructuring as a means for phonon engineering~[\onlinecite{cahill2003nanoscale,*cahill2014nanoscale,*volz2016}].  
	
\begin{figure}
	\includegraphics{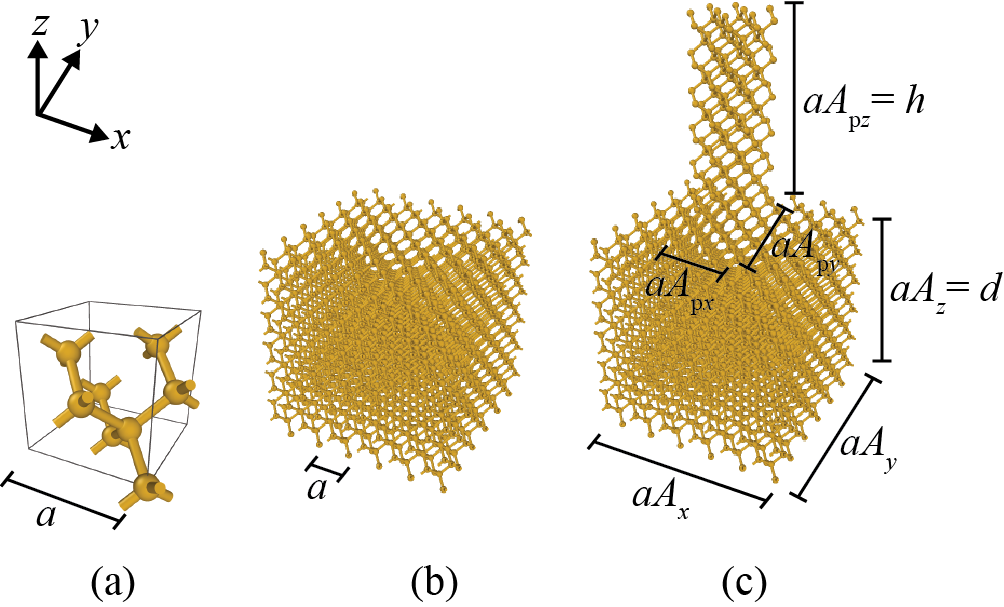} 
	\caption{\label{fig:fig_1} (a) Silicon conventional cell with the lattice constant $a=5.431~\textnormal{\AA}$, (b) uniform (unpillared) membrane unit cell, and (c) NPM (pillared) unit cell. The $x$-$y$ plane corresponds to the (001) plane of a silicon crystal.}
\end{figure}

\begin{figure}
 \includegraphics{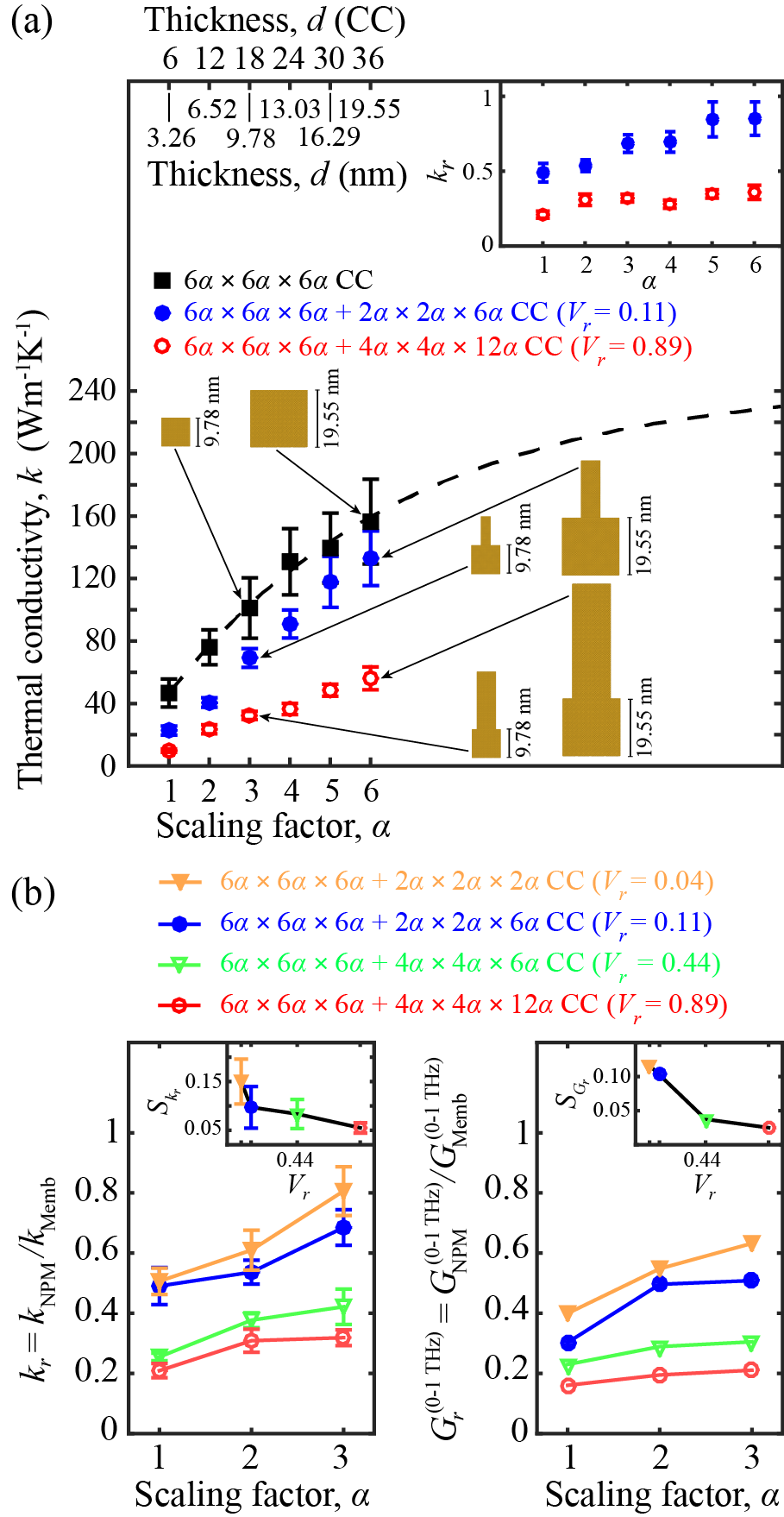} 
\caption{\label{fig:fig_2} (a) NPM and membrane thermal conductivity as a function of overall unit-cell size. NPM-to-membrane thermal conductivity ratio as a function of size is shown in the inset.~(b) Correlation between thermal conductivity reduction (left) and group velocity reduction (right) as a function of size for NPMs exhibiting different $V_r$ values. Central-difference slopes of the curves are plotted in the insets.}
\end{figure}

\indent Thermoelectric energy conversion stands to benefit profoundly from the ability to alter the phonon properties by nanostructuring~[\onlinecite{chen2003recent,*VineisAM2010}], as well as by reducing the material dimensionality~[\onlinecite{dresselhaus2007new}].~Thermoelectric materials, which generate electricity from heat and vice versa, are characterized by a figure of merit defined as $ZT = \sigma T S^{2} /k$, where $S$ is the Seebeck coefficient, $\sigma$ is the electrical conductivity, $k$ is the thermal conductivity (consisting of a lattice component and an electrical component), and $T$ is absolute temperature~[\onlinecite{Rowe2005}].~One strategy to improve the value of $ZT$, particularly in semiconductors, is to reduce the lattice thermal conductivity and attempt to do so without negatively affecting $S$ and $\sigma$.~A promising approach for achieving this goal is to introduce nanoscale local resonators as intrinsic substructures within, or attached to, a host crystalline material~[\onlinecite{PhysRevLett.112.055505},~\onlinecite{honarvar2016PRB}].~The emerging system, called \it nanophononic metamaterial \rm(NPM), exhibits unique properties that are not attainable in conventional nanostructured media such as nanocomposites~[\onlinecite{Liu2012nanocomp,*Kanatzidis2012}] or nanophononic crystals~[\onlinecite{Yu_2010,*Davis_2011}].~The substructure resonances, which could be numerous for relatively large substructures, may be tuned to couple with all or most of the heat-carrying phonon modes of the underlying host medium.~This atomic-scale coupling mechanism is essentially a \it resonance hybridization \rm between the wavenumber-dependent wave modes of the host medium (phonons) and the wavenumber-independent vibration modes of the local substructure (vibrons).~The outcome is significant reductions in the phonon group velocities across roughly the full spectrum of the host medium which, as a result, causes a lowering of the overall lattice thermal conductivity.~In Refs.~[\onlinecite{PhysRevLett.112.055505}] and~[\onlinecite{honarvar2016PRB}], the host medium is a free-standing silicon membrane and the resonators take the form of silicon nanopillars extruded from one surface.~This setup has the advantage that the thermal conductivity reducing elements (the nanopillars) are placed external to the main body of the membrane where in-plane heat transfer takes place, thus providing an environment for minimum electron scattering within the internal domain of the membrane.~Using kinetic theory and atomic-scale simulations, a factor of 2 reduction in the thermal conductivity was reported for very small unit cells with membranes less than 5 nm in thickness.~A recent molecular dynamics study examined a wider range of geometric dimensions on the same pillared silicon membrane configuration and demonstrated a reduction in the thermal conductivity by roughly a factor of 3~[\onlinecite{Wei2015JAP}].~The resonance-hybridization mechanism described above may be used in conjunction with boundary scattering from rough surfaces to lower $k$~[\onlinecite{Neogi_2015},\onlinecite{Neogi_2015_EJPB}].\\       
\indent For the concept of an NPM to be realized in practice, the nanostructured material's characteristic length scales, such as the membrane thickness and the nanopillar height and cross-sectional area, need to be at least on the order of a few tens of nanometers$-$which is substantially larger than the length scales considered in the previous studies mentioned above.~In this Letter, we study the effect of increasing size, in different ways, on the performance of an NPM. While we still consider relatively small models, we seek to establish the fundamental characteristics that govern the relationships between performance and size in order to provide predictive guidelines for much larger systems. We again consider a freestanding silicon membrane with a periodic array of silicon nanopillars standing on the surface$-$which is the same configuration studied in Refs.~[\onlinecite{PhysRevLett.112.055505}],~[\onlinecite{honarvar2016PRB}] and~[\onlinecite{Wei2015JAP}]. Specifically, we consider three size effects: nanopillar size (keeping the membrane thickness and lattice spacing constant), overall unit-cell size, and finite-structure size (which in practice is the sample size and is represented by the number of unit cells existing in a finite structure). \\ 
\indent For all our analysis, we consider a unit-cell model consisting of $A_{x}\times A_{y}\times A_{z}$ conventional cells (CC) of silicon forming the base (membrane portion) and $A_{\text{p}x}\times A_{\text{p}y}\times A_{\text{p}z}$ CC of silicon forming the nanopillar.~A silicon CC consists of an eight-atom cube with a side length of $a=0.5431$ nm (Fig.~\ref{fig:fig_1}a).~With this notation, the NPM unit cell has a membrane thickness of $d=aA_{z}$ and a nanopillar height of $h=aA_{\text{p}z}$.~For simplicity, we will limit our attention to nanopillars with a square cross-section, i.e., $b=aA_{\text{p}x}=aA_{\text{p}y}$.~For brevity, a labeling convention is adopted whereby the dimensions of a uniform (unpillared) unit cell are represented as $aA_{x}\times aA_{y}\times d$ (Fig.~\ref{fig:fig_1}b) and the dimensions of a pillared unit cell are represented as $aA_{x}\times aA_{y}\times d+b\times b\times h$ (Fig.~\ref{fig:fig_1}c). \\
\indent Equilibrium molecular dynamics (EMD) simulations are used in all studies, and non-equilibrium molecular dynamics (NEMD) simulations are additionally used for the finite-structure investigation.~\footnote{All MD simulations are performed using the LAMMPS software where the heat flux is evaluated based on a specified stress-based formula~[\onlinecite{plimpton1995fast}].}~For all simulations, room temperature, $T = 300$ K, is assumed and the Stillinger-Weber empirical potential is used to represent the interatomic interactions~[\onlinecite{stillinger1985computer}].~Furthermore, we only consider defect-free crystals.~In the EMD simulations, the computational domain consists of one unit cell and standard periodic boundary conditions are applied at the in-plane boundaries~[\onlinecite{honarvar2016PRB}].~EMD data are used within the Green-Kubo (GK) formalism~[\onlinecite{Zwanzig_1965,*Ladd_1986,*volz2000molecular},~\onlinecite{Schelling_2002}] to obtain the lattice thermal conductivity, which we will refer to as $k$ from now on for brevity.~\footnote{For the EMD simulations, the systems are initially equilibrated for 1 ns (with a time step ${\Delta}{t}=0.8$ fs) at the specified temperature using the $NPT$ ensemble (zero pressure cell size based on constant number of atoms, pressure and temperature).~Then, in the $NVE$ ensemble (constant number of atoms, volume and energy), the simulations are run for an additional 6 ns to collect heat fluxes that are recorded every 4 fs. The 6 ns time span is sufficiently large compared to the longest phonon lifetime.~With these parameters, the heat current auto-correlation functions converge within the first 500 ps.~The reported thermal conductivities are the average of values from six independent simulations with different initial velocities.~Furthermore, the thermal conductivities along both the $x$- and $y$-directions are considered, effectively resulting in an averaging over twelve predicted values.}~In contrast to EMD, NEMD predicts $k$ by direct application of the Fourier's law of heat conduction, defined as $k = -J/(A~\partial T/ \partial x)$ where $J$ is the heat current, $A$ is the cross section area (equal to $aA_{y}\times d$ or $aA_{x}\times d$), and $\partial T/ \partial x$ is the temperature gradient~[\onlinecite{Schelling_2002},~\onlinecite{Tenenbaum_1982},~\onlinecite{sellan2010size}].~A finite number of unit cells is used in the NEMD simulations.~Denoting $N_{x}$ and $N_{y}$ as the number of NPM unit cells in the $x$- and $y$-directions, respectively, we consider $k$ vs. $N_{x}$ while keeping $N_{y}=1$. Periodic boundary conditions are applied along the $y$-direction and Langevin heat baths are used to apply a temperature gradient across the $x$-direction where the left-end and right-end temperatures are set to $T_{L}=310$ K and $T_{R}=290$ K, respectively.~\footnote{The NEMD simulations are run using ${\Delta}{t}=0.8$ fs for 0.8 ns to reach the steady state and then for another 9.6 ns to obtain an average heat flux and temperature profile; the reported thermal conductivities are the average of two independent simulations.}\\
\indent We commence by examining the effect of the nanopillar size.~Two NPM unit cells are considered with dimensions $6\times 6\times 6+2\times 2\times 6$ CC and $6\times 6\times 6+4\times 4\times 12$ CC, respectively. The nanopillar-to-base membrane volume fraction is $V_r=0.11$ for the unit cell with a small nanopillar and $V_r=0.89$ for the unit cell with a large nanopillar.~The corresponding thermal conductivity ratio $k_r$ (defined as the NPM thermal conductivity, $k_{\rm NPM}$, divided by the thermal conductivity of a corresponding uniform membrane sized $6\times 6\times 6$ CC, $k_{\rm Memb}$) is 0.49 and 0.2, respectively.~This indicates that the intensity of thermal conductivity reduction increases with $V_r$, which is expected because a larger nanopillar exhibits more local resonance modes than a smaller one and therefore yields a stronger effect for a given membrane thickness and nanopillar lattice spacing. More local resonances implies more mode hybridizations and consequently a more intense overall group velocity reduction.~It is noteworthy that a $k_r=0.2$ value corresponds to a factor of 5 reduction in the thermal conductivity, which is the highest to date for an NPM. We also observe that $k_r$ versus $V_r$ drops with a decreasing rate; this is because as the number of resonances increases the hybridization effect eventually reaches saturation.\\
\indent We now investigate the effect of unit-cell size for each of the configurations considered above.~This is done by proportionally expanding all dimensions of the NPM (and corresponding uniform-membrane) unit cells and tracking the thermal conductivity reduction with this uniform increase in size.~For the $V_r=0.11$ and $V_r=0.89$ cases, we respectively consider unit-cell dimensions $6\alpha\times 6\alpha\times 6\alpha+2\alpha\times 2\alpha\times 6\alpha$ CC and $6\alpha\times 6\alpha\times 6\alpha+4\alpha\times 4\alpha\times 12\alpha$ CC for the following scaling factor values: $\alpha =$ 1, 2, 3, 4, 5 and 6.~This set covers values of membrane thickness ranging from $d=$ 3.26 nm ($\alpha =$ 1) to $d=$ 19.55 nm ($\alpha =$ 6), advancing in increments of $\Delta d=$ 3.26 nm.  \\
\indent The unit-cell size effect results are shown in Fig.~\ref{fig:fig_2}a where $k_{\rm NPM}$ for each of the two cases is plotted against $\alpha$ (and $d$).~The thermal conductivity for a corresponding uniform membrane is also plotted to serve as the reference case.~\footnote{All uniform membrane calculations are based on  $6\times 6\times d$ CC unit cells, for which the thermal conductivities are converged as a function of the simulation cell size~[\onlinecite{honarvar2016PRB}].}~The resulting $k_r$ values are given in the inset. We observe, expectedly, $k_{\rm Memb}$ to drop significantly with decreasing $d$ due to dispersion modification as well as the reduction of phonon mean free paths (MFP) caused by diffuse boundary scattering at the surfaces as a result of their reconstruction at the equilibrium state~[\onlinecite{Appclbaum_1976,*Gomes_2006}].~These effects been studied experimentally in the literature in the context of a thin silicon layer on a substrate~[\onlinecite{Asheghi_1997,*Marconnet_2013}], freestanding silicon membranes~[\onlinecite{Neogi_2015},\onlinecite{Cuffe_2015}], and also silicon nanowires~[\onlinecite{Li_2003}].~The $k_{\rm NPM}$ curves are observed to similarly increase with increasing unit-cell size (due to increasing thickness) until gradual saturation.~It is seen, however, that the thermal conductivity of the NPM with large nanopillars grows with unit-cell size at a lower pace than the NPM with small nanopillars.~This behavior is desirable as it suggests that a higher starting $V_r$ value leads to a less negative effect of increasing size.~These trends will continue until the unit-cell characteristic length scales exceed the full span of the MFP distribution$-$at this point, nonlinear scattering mechanisms will dominate and the likelihood of occurrences of resonance hybridizations will diminish as a result.\\        
\begin{figure}[t!]
	\includegraphics{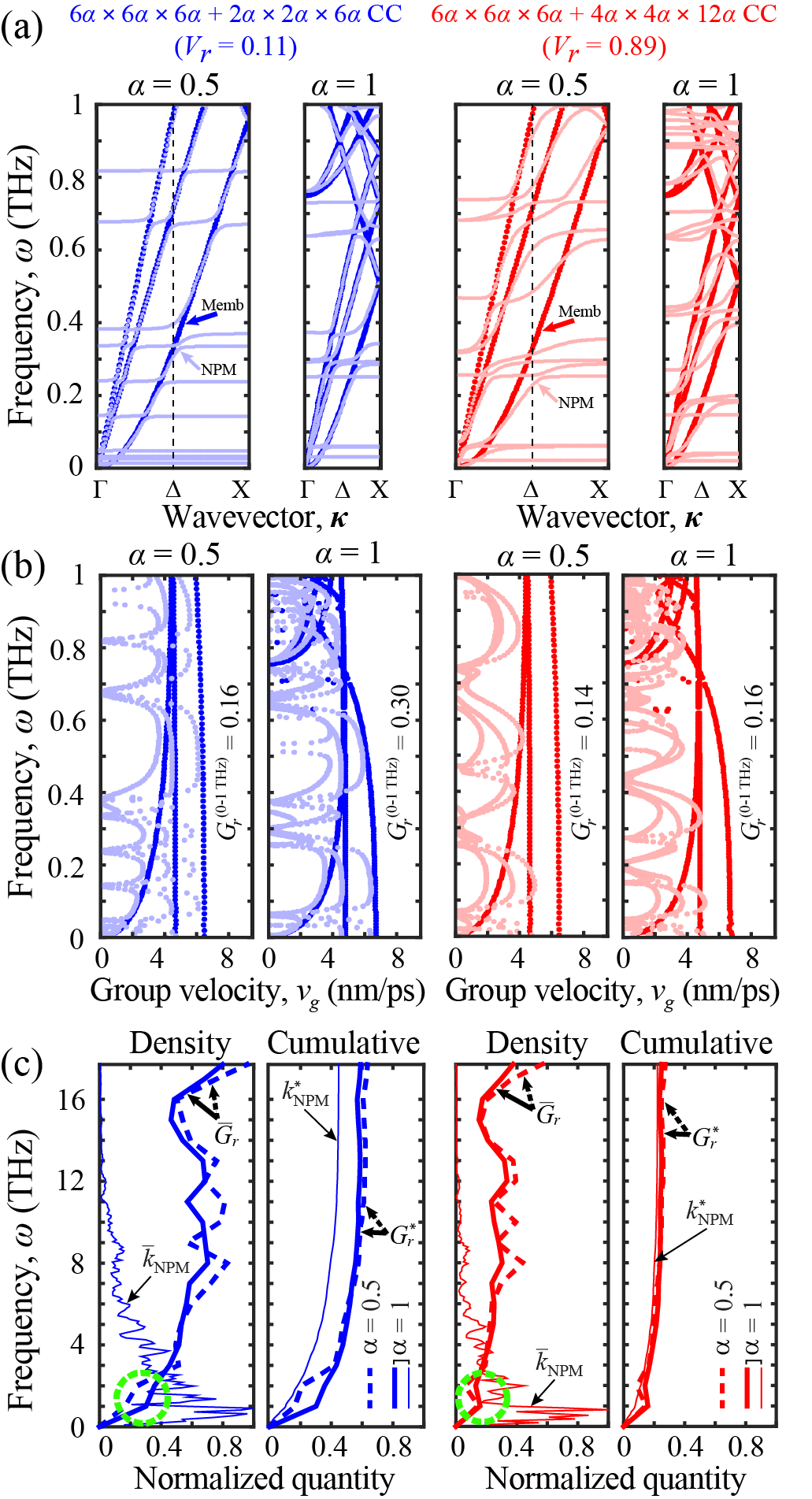} 
	\caption{\label{fig:fig_3} (a) Dispersion curves and (b) group velocities for $V_r=0.11$ and $V_r=0.89$ NPMs for two different unit-cell sizes.~The membrane and NPM quantities are represented by dark and light colors, respectively.~(c) Corresponding density, $\bar{G}_{r}$, and cumulative, ${G}_{r}^{*}$, quantities of the ratio of NPM-to-membrane group velocities: $\bar{G}_{r}$ is $G_{r}^{(\lambda_2-\lambda_1)}$ evaluated for incremental $\Delta \lambda=\lambda_2-\lambda_1=1$ THz windows, and ${G}_{r}^{*}$ is $G_{r}^{(\lambda_2-\lambda_1)}$ evaluated for a cumulative $\Delta \lambda=\lambda_2-0$ window.~The dashed and solid lines correspond to $\alpha=0.5$ and $\alpha=1$ unit-cell sizes, respectively.~Density, $\bar{k}_{\rm NPM}$, and cumulative, ${k}_{\rm NPM}^{*}$, thermal conductivity quantities for $\alpha=1$ are plotted in the background: $\bar{k}_{\rm NPM}$ is ${k}_{\rm NPM}(\omega)$ normalized with respect to its maximum value, and ${k}_{\rm NPM}^{*}$ is $\int_{0}^{\omega} {k}_{\rm NPM}(\omega') d\omega'$ normalized with respect to ${k}_{\rm Memb}$. Low-frequency behavior is highlighted.}. 
\end{figure}
\begin{figure}[t!]
	\includegraphics{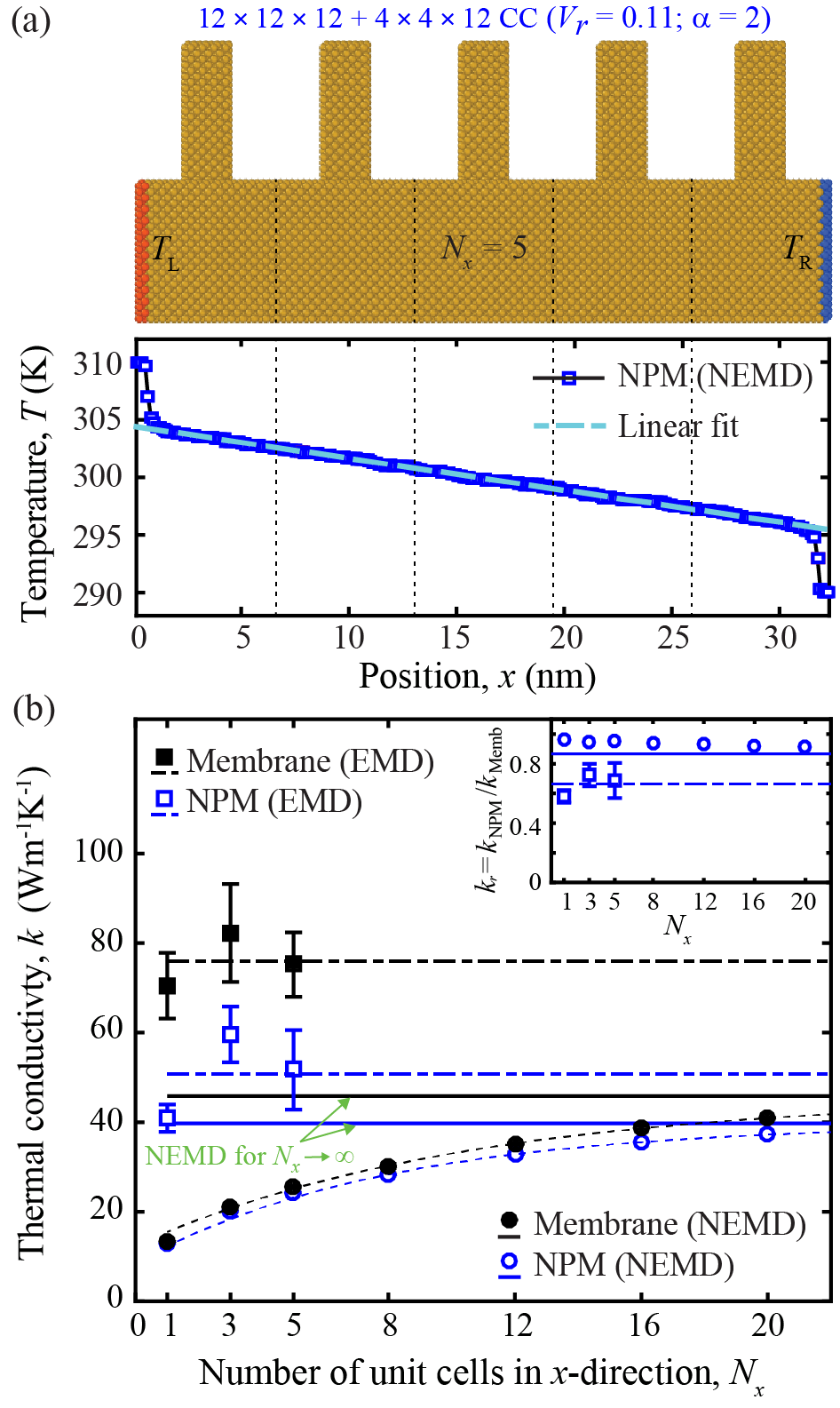} 
	\caption{\label{fig:fig_4} (a) NEMD simulation setup ($N_x=5$ in this schematic) and temperature profile across the finite dimension.~(b) Thermal conductivity as a function of sample size, shown in normalized form in the inset.~The curved dashed lines show exponential fittings.~The horizontal lines represent average values for the EMD results (dashed) and converged values for the NEMD results (solid).}
\end{figure}
\indent The observed $V_r$ and $\alpha$ dependencies in the reductions of the thermal conductivity values are explained by comparing with the corresponding phonon group velocity trends.~We conduct lattice dynamics (LD) calculations~[\onlinecite{gale2003general}] and obtain phonon group velocities for the models considered, as well as two more $V_r$ values, up to $\alpha=3$.~For efficient calculations, the dispersion spectrum is computed for only a low frequency range, $0\le \omega \le 1$ THz, using the reduced Bloch mode expansion technique (RBME)~[\onlinecite{Hussein_PRSA_2009}].~\footnote{RBME is a secondary modal-expansion technique for speeding up band-structure calculations. In the present implementation, a 3-point expansion is conducted whereby eigenvectors are selected at the $\Gamma$,~$\Delta$ and X points within the irreducible Brillouin zone to form the reduced basis.~This procedure generates dispersion curves at more than an order of magnitude higher speed with errors less than 1\% compared to full (non-reduced) calculations.}~The group velocities, $v_g(\bm{\kappa}, m)$, where each phonon mode is labeled by wavevector $\bm{\kappa}$ and mode number $m$, are obtained by differentiating the frequency dispersion curves over 129 wavevector points by a 3-point finite-difference scheme.~An average group velocity quantity is then obtained for each case, defined as ${G^{(\lambda_1{\text{-}}\lambda_2)}} = [1/(n_{\kappa} n_{\lambda})] \textstyle\sum_{\kappa}^{n_{\kappa}} \textstyle\sum_{m}^{n_{\lambda}} v_g(\kappa, m)$ where ${n_{\kappa}}$ is the number of wavevector points and ${n_{\lambda}}$ is the number of modes between frequencies $\lambda_1$ and $\lambda_2$.~In Fig.~\ref{fig:fig_2}b, we plot $k_r$ versus $\alpha$ and the ratio of the NPM-to-membrane average group velocity also versus $\alpha$.~The latter quantity is denoted $G_{r}^{(\lambda_1-\lambda_2)}=G_{\rm NPM}^{(\lambda_1-\lambda_2)}/G_{\rm Memb}^{(\lambda_1-\lambda_2)}$ and is evaluated considering all group velocities from $\lambda_1=0$ to $\lambda_2=1$~THz.~The central-difference slopes of the two sets of curves are shown in the insets.~We see a clear qualitative correlation between the $k_r$ and $G_{r}^{(0{\textbf{-}}1 {\text{THz}})}$ curves which indicates that the thermal conductivity trends are driven by the manner by which the resonance hybridizations affect the group velocities for the different values of $V_r$ and $\alpha$.~Of particular importance is the frequency distribution of the nanopillar vibration modes with respect to the underlying membrane dispersion curves for the various cases.~This comparison is demonstrated more explicitly in Fig.~\ref{fig:fig_3}, where we show the dispersion and raw group velocity curves for the $V_r=0.11$ and $V_r=0.89$ cases for unit-cell sizes $\alpha=0.5$ and $\alpha=1$.~For example, we observe that a sharp contrast between the NPM and membrane group velocities extends over a broader frequency range for $\alpha=0.5$ compared to $\alpha=1$ which explains why $k_r$ rises with $\alpha$.~As for the slopes in Fig.~\ref{fig:fig_2}b, these drop with $V_r$ because as the number of resonances increases the hybridization-induced changes with size eventually reaches saturation.~Figure~\ref{fig:fig_3}c presents the group-velocity density and cumulative values highlighting strong reductions due to the presence of the nanopillars in the low-frequency range of $0-3$ THz, which as demonstrated by the overlaying thermal conductivity curves~\footnote{The frequency-dependent thermal conductivity curves are obtained using the Boltzmann transport equation following the single-mode relaxation time approximation as detailed in Ref.~[\onlinecite{PhysRevLett.112.055505}]} is the most dominant range contributing to the overall value of $k_{\rm NPM}$.~We note that the low-frequency dominance and the monotonic increases of the difference in the cumulative group velocity curves between the small and large unit-cell sizes up to roughly 3 THz justifies our choices of $\lambda_1=0$ and $\lambda_2=1$~THz for the comparative analysis presented in Fig.~\ref{fig:fig_2}b.\\
\indent Finally, we use NEMD simulations to investigate the dependency of the thermal conductivity on the size of a finite structure, which is illustrated in Fig.~\ref{fig:fig_4}.~As the number of unit cells $N_{x}$ increases, phonons with longer wavelengths become available for carrying the heat and thus the thermal conductivities for the uniform and NPM cases gradually increase until they converge to their large sample size values; see, for example, Refs.~[\onlinecite{Schelling_2002},~\onlinecite{Tenenbaum_1982},~\onlinecite{sellan2010size}] for length-dependency studies on other material systems.~We observe that the difference between the two $k$ curves increases with $N_{x}$, although slightly$-$possibly this is because more of the long-wave phonons with relatively high group velocities become available for resonance hybridization as the number of unit cells is increased.~That difference is also expected to level off for large $N_{x}$.~For comparison, corresponding EMD results are shown.~While both indicate NPM thermal conductivity reduction, the EMD predictions are higher than their counterparts from the NEMD simulations.~This discrepancy between the two sets of predictions is not uncommon because the EMD and NEMD methods depend on different factors for their accuracy~[\onlinecite{sellan2010size},~\onlinecite{Landry2008}].\\   
\indent In summary, we have established that the extent of thermal conductivity reduction by resonance hybridization increases with the size of the local resonator but decreases with increased overall unit-cell size. However, the rate of this decrease in performance with unit-cell size is inversely proportional to the volumetric ratio of the resonator (nanopillar) to the host medium (base membrane).~Thus an NPM with a high value of $V_r$ will sustain most of its performance in large sizes.~We have also shown that sample size has a mild effect on the thermal conductivity reduction.~The NPM unit cells investigated range in base-membrane thickness from 3.26 to 19.55 nm.~One of the pillared silicon membrane configurations considered exhibit a factor of 5 thermal conductivity reduction compared to a corresponding uniform membrane.\\
\indent This research was supported by the National Science Foundation (USA) CAREER Grant No.~1254931. The authors are most grateful to Prof.~Pierre Deymier and Dr.~Nick Swinteck for their assistance with the initial set up and usage of the LAMMPS code for the equilibrium MD simulations. This work utilized the Janus supercomputer, which is supported by the National Science Foundation (USA) (Award No. CNS-0821794) and the University of Colorado Boulder. The Janus supercomputer is a joint effort of the University of Colorado Boulder, the University of Colorado Denver, and the National Center for Atmospheric Research (USA).	
\bibliography{RefsSize}

\end{document}